# A generalized software framework for consolidation of radiotherapy planning and delivery data from diverse data sources.

Yasin Abdulkadir[1], Justin Hink[1], Peter Boyle[1], Dishane Luximon[1], Justin Pijanowski[1], Timothy Ritter[2], Bruce Curran[2], Min Leu[3], Nicholas Nickols[3], Steve P. Lee[4], Jatinder R. Palta[2], Maria Kelly[5], Rishabh Kapoor[2], Reid F. Thompson[6], Daniel A. Low[1], James M. Lamb[1], Jack Neylon[1]

1. Department of Radiation Oncology, University of California, Los Angeles
2. Virginia Commonwealth University Health System
3. VA Greater Los Angeles Healthcare System
4. VA Long Beach Healthcare System
5. Department of Veterans Affairs National Radiation Oncology Program
6. VA Portland Healthcare System

**Abstract**

Aggregating large-scale radiotherapy planning and delivery data is crucial for advancing radiation oncology research and improving clinical practice, yet challenges persist due to the diversity of treatment planning systems (TPS), record and verify (R&V) systems, and complex data formats lacking standardized retrieval methods. We developed a robust software framework that automates the collection and integration of multi-institutional radiotherapy data from diverse TPS and R&V systems. By utilizing the unidirectional references of DICOM objects, our framework reconstructs complete patient datasets starting from Radiotherapy Treatment Records (RTRECORDs), managing tasks such as data queries, transfers, verification, and logging. It effectively maps DICOM linkages between RTRECORDs, RTPLANs, RTDOSEs, RTSTRUCTs, planning images, registrations, and associated diagnostic images, incorporating custom modules for data conversion and comprehensive error handling. Implemented across multiple institutions using various systems—including ARIA, Eclipse, MOSAIQ, RayStation, MIM, and Pinnacle—the framework successfully collected data from two clinics over an 11-year period, aggregating data from 6,022 patients and 13,871 treatment plans with a success rate of 99.76% and an average processing time of approximately 18 minutes per patient. Ongoing efforts are extending data collection to clinics lacking DICOM Query/Retrieve capabilities, demonstrating the framework's adaptability to various clinical environments. This efficient automation of comprehensive data collection overcomes significant technical barriers, facilitating the creation of large-scale datasets that can accelerate advancements in radiation oncology.

## 1. Introduction

Very large databases of radiotherapy planning and delivery data have immense potential to advance radiation oncology research and clinical practice. Such "radiotherapy big data" can facilitate the development and validation of predictive models, enhancing outcomes and treatment efficiency[1], and can reveal patterns and correlations that smaller datasets cannot, thereby improving the precision and personalization of radiotherapy treatments[2]. Additionally, radiotherapy big data may enable artificial intelligence based automated segmentation and radiotherapy planning, allowing improved quality, safety and efficiency of treatments, and potentially leveraging the expertise of tertiary providers to under-resourced clinics that treat underserved patients.

Several initiatives have made significant strides in the retrospective collection of radiotherapy data. The Cancer Imaging Archive (TCIA), for instance, hosts a vast array of imaging datasets, including those in DICOM-RT format, demonstrating the feasibility and utility of aggregating radiotherapy data from diverse sources. TCIA relies on contributing sites to meticulously curate and submit data, ensuring its consistency and usability[3]. However, as of June 16th, 2024, TCIA contains only ~12,000 radiotherapy patient datasets with labeled anatomy (i.e., RTSTRUCT) and ~2,000 patient datasets with radiation treatment plan dose (RTDOSE). Another notable initiative is the National Radiation Oncology Registry (NROR), which was established to collect and analyze radiation treatment data on a large scale, promoting evidence-based practices[4]. The University of Michigan's Radiation Oncology Analytics Resource (M-ROAR) exemplifies a successful effort in aggregating key multidisciplinary data elements, supporting clinical and research applications with data for over 17,000 patients treated since 2002[3]. The Veterans Health Administration's HINGE platform exemplifies an advanced, automated system for quality surveillance and outcome assessment in radiation oncology, integrating data from treatment planning, management, and electronic health records to facilitate quality improvement, predictive analytics, and enhanced clinical decision-making[5].

Despite the progress made thus far, we and others believe that, yet larger databases would provide increased benefit to radiotherapy research. For example, the creation of ImageNet[6], containing 10M manually labeled images, was instrumental to the development of image classification algorithms receiving widespread public usage. However, the technical challenges in creating radiotherapy big data enterprises are considerable. Assembling such data requires merging multiple disparate clinical data sources. Prospective data collection allows prospective definition of protocols and data formats, facilitating collection and storage, but the extensive time required, potentially spanning over a decade, delays potential advancements and benefits. No single institution treats enough patients to reach the desired data threshold within a reasonable timeframe, necessitating a collaborative effort. This collaboration is complicated by the need to agree on and enforce common data formats across institutions. Prospective data collection also faces logistical and financial barriers associated with obtaining informed consent from patients, as managing informed consent on a large scale is nearly insurmountable. Institutional Review Boards (IRBs) often require informed consent for prospective data collection, hindering the functionality of proposed systems[4]. Given these constraints, retrospective data collection presents itself as a viable option. However, retrospective data collection within a single institution must contend with the diversity of data sources and formats.

Retrospective data collection from multiple institutions compounds the complexity, necessitating effective strategies for standardizing and integrating data from multiple sources with differing formats into a cohesive database. Ensuring data interoperability requires robust frameworks for data standardization and normalization. Furthermore, the sheer volume of data necessitates advanced storage solutions and efficient data management systems capable of handling large-scale datasets[7]. Maintaining data quality and completeness is crucial, as inconsistencies and gaps in data can lead to erroneous analysis and conclusions. Therefore, implementing automated data validation and correction mechanisms is essential. Ensuring patient privacy and data security is also paramount, given the sensitive nature of medical data. Compliance with regulations such as HIPAA requires stringent security protocols and anonymization techniques to protect patient information[8].

Here, we describe the architecture and report the initial performance of a software framework designed for the automated collection of multi-institutional planning data stored in diverse treatment planning systems (TPS) and record and verify (R&V) systems. The framework is built to operate efficiently, ensuring timely data collection and processing. It includes mechanisms to handle errors and interruptions gracefully, maintaining continuous operation. Additionally, it verifies the delivery of radiotherapy treatment plans to exclude incomplete or irrelevant data (such as unused trial plans), interfaces with multiple planning systems including those without DICOM interfaces and manages diverse data formats and sources to ensure comprehensive data integration. This robust and versatile framework aims to streamline the aggregation of radiotherapy data, thus facilitating large-scale analyses and fostering advancements in treatment quality and safety.

## 2. Methods

### 2.1 Data Collection and Framework Structure

Each DICOM object has a unique identifier (UID). References to these UIDs create linkages between objects. However, these references are rarely reciprocal, limiting recreation of the DICOM data structure for a patient to a unidirectional path. The DICOM structure followed in the code involves several key components:

- Radiotherapy Treatment Record (RTRECORD): The starting point for data collection, containing references to the Radiotherapy Treatment Plan (RTPLAN).
- RTPLAN: Linked from RTRECORD, it contains treatment planning information.
- Radiotherapy Dose (RTDOSE): Connected to RTPLAN, it contains dose distribution data.
- Radiotherapy Structure Set (RTSTRUCT): Linked from RTPLAN, it contains structure set data (contours).
- Planning CT/MR: Connected to RTSTRUCT, it contains the planning imaging data.

- Registration (REG): Registration data that links planning images to other modalities like CT, MR, or PET.
- Registered CT/MR/PET: Imaging data registered to the planning images.

The logical processing algorithm for this framework is referred to as the 'CORE', which centrally manages tasks like data queries, transfers, verification, and logging. The linear steps taken to map the DICOM linkages and gather the objects from a DICOM-compliant interface are as follows, starting at the point of a successful query of an RTRECORD from a R&V system:

1. Move the RTPLAN referenced by returned RTRECORDs to the CORE, by UID.
2. Move the RTSTRUCT referenced by the RTPLAN to the CORE, by UID.
3. Query/Retrieve (Q/R) the RTDOSE that references the RTPLAN.
4. Q/R all RTRECORDs referencing the RTPLAN.
5. Q/R the CT SIM slice that is referenced by the first slice of the first non-empty structure in the RTSTRUCT.
6. Q/R the CT SIM series to which the slice identified in Step 5 belongs.
7. Q/R the REG that references the CT SIM slice.
8. Q/R for all CT/MR/PET series that share the same Frame of Reference UID with the CT SIM.
9. Q/R the diagnostic images referenced by the REGs that reference the CT SIM.
10. Q/R all images that share a Frame of Reference UID with images identified in Step 9.

**Figure 1** presents a graphic depiction of DICOM objects representing radiotherapy planning and delivery, the linkages between them, and the steps outlined above to map the linkages between the objects. Once each object was identified, it was queued for DICOM transfer to a PACS (Picture Archiving and Communication System) that we set up. Prior to moving an object, it was verified through Q/R, that it was not already in the PACS. Our primary objective was to gather data corresponding only to treated radiotherapy plans, making the RTRECORD the starting point for assembling a patient dataset. By initiating the data collection with the RTRECORD, we ensured that the dataset was anchored in the actual treatments administered. To streamline this process, we developed methods to inspect DICOM reference connections, which facilitated the identification and linkage of relevant data objects. This approach minimized unnecessary data transfers by ensuring that only essential and directly connected data were collected. However, there is added practical complexity because intuitive concepts of belongingness are not always reflected in DICOM references. For example, in DICOM, the RTPLAN object does not contain a pointer to an RTDOSE object; rather, the RTDOSE object contains a pointer to the RTPLAN object (its SOPInstanceUID). Therefore, in order to find the RTDOSE that corresponds to RTPLAN, it was necessary to search through all the RTDOSE objects with the same StudyInstanceUID (the parent object that both RTPLAN and RTDOSE belong to) as the RTPLAN and determine which had a pointer to the RTPLAN. These DICOM reference relationships are represented as red arrows in **Figure 1**. Finally, in clinical treatment planning software, when registering a PET/CT to a simulation CT, an explicit REG object is created for the CT-CT registration, but the PET-CT registration is based on the shared frame of reference meaning that the PET and CT images acquired during the PET/CT scan inherently share the same coordinate system. This shared frame of reference allows the PET and CT images to be automatically aligned without the need for additional registration processes between them. Thus, when attempting to gather all diagnostic images registered to the simulation CT, it is necessary to obtain the diagnostic images that are referenced by REG objects, as well as all diagnostic images that share a FrameOfReferenceUID with explicitly referenced images.

## 2.2 Data Sources

Treatment planning and delivery data was collected from multiple institutions employing various TPS and R&V systems, including ARIA and Eclipse (Siemens Healthineers, Erlangen, Germany), MOSAIQ (Elekta, Inc, Stockholm, Sweden), RayStation (Raysearch Laboratories, Stockholm, Sweden), MIM (GE Healthcare), and Pinnacle (Phillips). Where possible, vendor supported DICOM interfaces were used (ARIA/Eclipse, MIM, RayStation). Explicit conversion to DICOM was required for data from Pinnacle and MOSAIQ installations. **Figure 2** shows a diagram of data sources and how they are interfaced to program modules and mapped to the archival data repository.

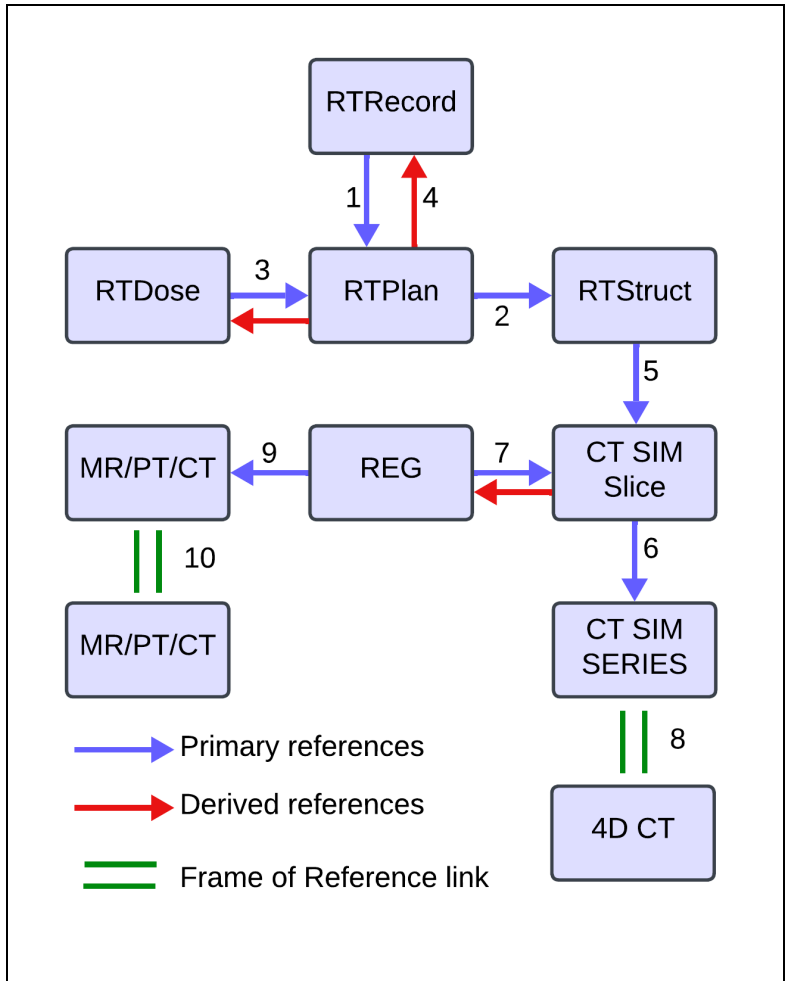

**Figure 1:** DICOM-RT connectivity used and defined by our software framework. Blue lines indicate native references stored within DICOM-RT files. Red lines indicate references derived by our code (i.e. by reverse lookup). Numerals indicate order of steps within the process to reconstruct the entire object chain (as described in **Section 2.1**)

## 2.3 Software Framework Architecture

The software framework developed for this study incorporates several critical components designed to ensure efficient and robust data collection from multiple sources. The architecture includes the following components:

<u>Job-Queuing System:</u> Central to the architecture, this system manages the collection process. Each job represents a DICOM object type and a corresponding set of actions. A Job Queue stores and processes jobs, ensuring efficient data handling and minimizing data transfers by inspecting DICOM reference connections. **Figure 3** diagrams the job-queuing system, its functionality, and its relationship to DICOM sources.

<u>Error Handling and Queue Recovery:</u> Error handling is necessary because individual DICOM commands sometimes fail due to network traffic, overloading of the called application entity, and other reasons. Our approach was that if a DICOM command returned an error code, then the job was re-instantiated on the Job Queue. To prevent infinite loops, each job had a counter that allowed only up to 10 attempts. To allow recovery from system crashes, the state of the Job Queue was written to file at configurable intervals (e.g. 1-minute intervals).

<u>Non-DICOM queryable data transfers:</u> The MOSAIQ R&V, Pinnacle TPS, and RayStation TPS do not provide DICOM queryable transfer interfaces. In a typical Pinnacle installation, most saved data is stored as proprietary formats bundled as "tar" files. RayStation allows scripted export of DICOM format plans but does not support DICOM query-retrieve. Our approach was to write dedicated code and processes to send these data (converting to DICOM if needed) to what we term an *Auxiliary PACS,* which then was interfaced to our main Job Queuing System. This was done in order to provide a single framework that connected treatment plan objects with the information about which plans were treated. Keeping in mind the overall goal to store only treated plans and retain the information on the number of fractions that each plan was treated.

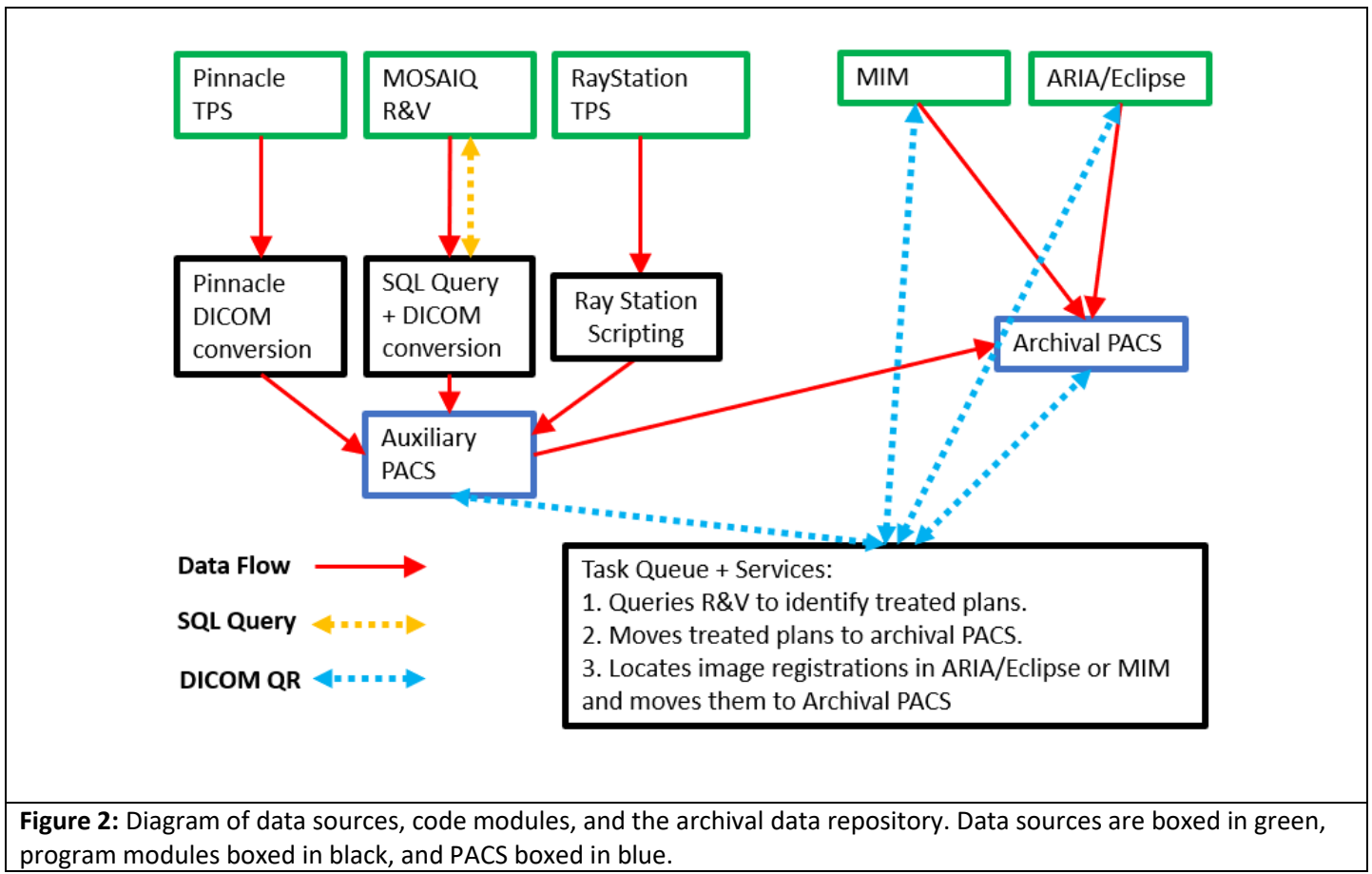

**Figure 2:** Diagram of data sources, code modules, and the archival data repository. Data sources are boxed in green, program modules boxed in black, and PACS boxed in blue.

Mosaiq DICOM Conversion Module: MOSAIQ provides treatment data (analogous to what is found in an RTRECORD) via SQL interface. DICOM RTRECORDs are not provided without the DataDirector purchasable option, which many clinics have not purchased. Thus, a software module was written to extract treatment information via SQL and write RTRECORDs.

Pinnacle tar file conversion: Conversion of Pinnacle tar files was performed by open-source code from the PyMedPhys project[9], which was further modified by our team for this application. Our enhancements enabled the processing of multiple patients within a single tar file, as well as bulk processing of all tar files in a directory. We ensured the generation of DICOM-compliant Unique Identifiers (UIDs) while maintaining correct reference relationships among the data. Additionally, we modified hard-coded tags to be automatically collected and implemented mechanisms to gracefully handle errors, logging them for manual processing.

MIM *Sessions* Data: MIM allows the possibility to store radiotherapy structure sets and image registrations in a *Sessions* format. While in the MIM Patient List User Interface (UI), structure sets and registrations appear to be stored as RTSTRUCT and REG objects, internally they are stored as DICOM RAW objects. Within the RAW objects, private tags hold DICOM header information identical to that found in RTSTRUCT and REG headers. Thus, we developed code to unpack the Sessions and convert the contents to RTSTRUCT and REG objects.

Unit and Integration Testing: Pytest[10] was used for both unit and integration tests. Extensive tests were conducted on individual modules to ensure each component operates correctly. Integration tests were conducted to verify that the entire system functions as expected when all components are combined.

Error Handling and logging: Error handling ensures robust operation despite potential issues during data collection and processing. Logging provides detailed documentation of all activities related to data extraction, transformation, and storage and allows for efficient recovery from crashes due to network or hardware issues (e.g. computer was turned off) that were inevitable over multiple weeks of running. Comprehensive error handling routines are incorporated, capable of identifying, logging, and recovering from errors. These routines include retry mechanisms and alerts for manual intervention when necessary. Logging mechanisms are integrated within the Python SCU, Python SCP, and Orthanc components. Logs include detailed information about data transfer activities, errors, and recovery actions. Logging was implemented by writing to an SQLite database; log reading was performed using SQL queries. All software modules logged errors into the same database. This allowed faster and more robust searching of the extensive log files consisting of millions of logged messages.

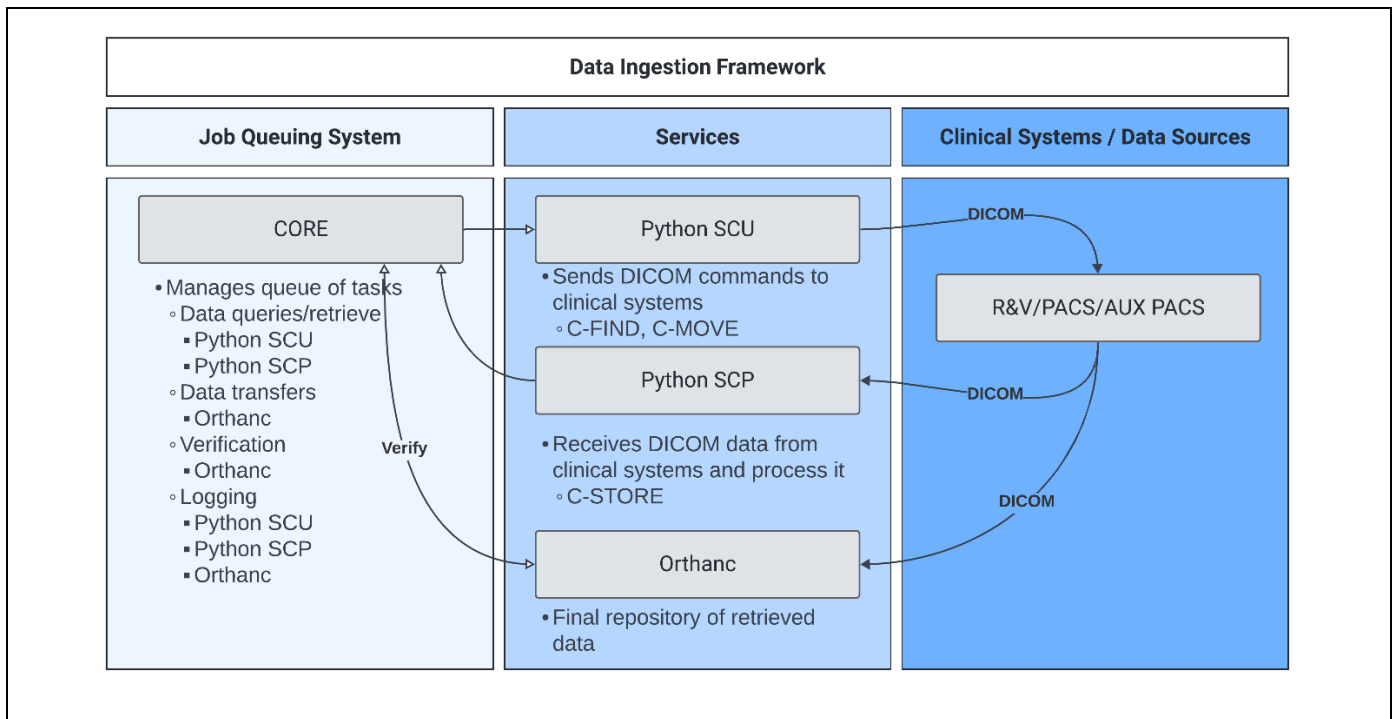

**Figure 3**: Data Ingestion Workflow. The 'CORE' centrally manages jobs like data queries, transfers, verification, and logging. The Python SCU communicates with clinical systems using DICOM commands (C-FIND, C-MOVE), while the Python SCP handles incoming DICOM data via C-STORE and generates jobs for the queue. 'Orthanc' serves as the final repository, ensuring data integrity and accessibility.

Data Storage and Management: All data, from whatever source, was required to be ultimately aggregated into a single, cohesive database. The Orthanc open-source PACS[11,12] was selected as such destination. Orthanc was chosen for several reasons. First, we felt it was essential that the data was stored in a PACS to facilitate further processing and/or transfer to larger downstream database, because a PACS is queryable and manages duplicate data storage requests transparently. Secondly, Orthanc has an integrated SQL database that allows for the storage of ancillary information on a per-object basis in a way that is transparent to the end user (e.g. we chose to add information on the provenance of each DICOM object), and also supports the REST API which allows our custom software to make calls directly to the underlying Orthanc program (i.e. at a more direct level than performing DICOM queries).

Data Validation and Manifest Generation Program: A software module was created to validate collected data by performing error and consistency checking, and to generate human-readable summaries. This program module inspects DICOM reference connections to identify missing data. It provides summaries to facilitate human review and ensure data completeness.

Data and Code Availability: The software code developed during this project will be made publicly available upon its completion to support transparency and facilitate further research in the field. However, the primary radiotherapy data collected are not publicly available due to patient privacy concerns and regulations under the Health Insurance

Portability and Accountability Act (HIPAA). Access to this data is restricted to protect patient confidentiality, and any sharing of de-identified data will be considered in compliance with relevant privacy laws and institutional policies.

**3. Results**

Variability of Clinical Systems

To better understand the scope and complexity of retrospective data extraction within the Veterans Affairs (VA) health system, we first sought to explore the archival data landscape across VA clinics. This exploration aimed to identify the challenges associated with aggregating radiotherapy data from diverse clinical systems, which is crucial for developing an effective data collection framework.

In a survey of nearly 40 clinics, only 31% reported a single data repository for all their radiotherapy data (planning and treatment) from the past decade. Multiple R&V systems were reported at 28% of clinics, multiple TPS were reported at 46%, and 44% had a separate system for performing image fusions outside of their TPS.

**Figure 4(a)** displays the percentage of clinics with all treatment data in R&V systems that natively support DICOM Q/R, while **Figure 4(b)** displays the percentage of self-reported estimates of plans stored in TPS that natively support DICOM Q/R.

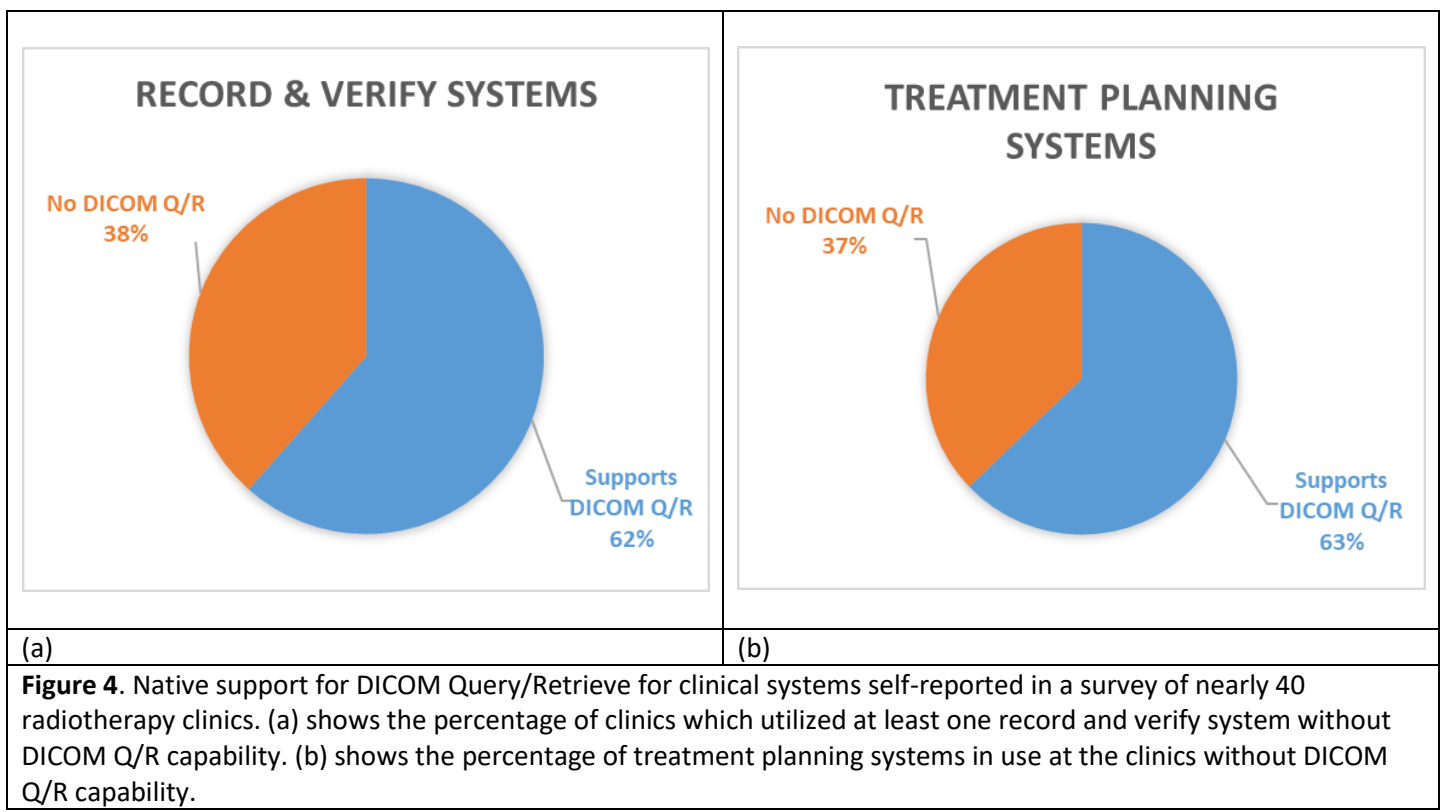

(a) (b)
**Figure 4**. Native support for DICOM Query/Retrieve for clinical systems self-reported in a survey of nearly 40 radiotherapy clinics. (a) shows the percentage of clinics which utilized at least one record and verify system without DICOM Q/R capability. (b) shows the percentage of treatment planning systems in use at the clinics without DICOM Q/R capability.

Data Collection Overview

To transition from understanding the broader challenges of data aggregation to testing our proposed solution, we selected two clinics from the surveyed sites for initial implementation of our data ingestion framework. These clinics were chosen based on specific criteria identified in our earlier exploration of the archival data landscape. Specifically, we targeted clinics where nearly all radiotherapy data was stored in systems that natively support DICOM Query/Retrieve (Q/R), as highlighted by the 31% of clinics with a single data repository (**Figure 4**). This selection allowed us to focus on assessing the framework's performance in an environment with fewer technical barriers, thereby establishing a baseline before scaling up to more complex settings.

We successfully collected radiotherapy planning and treatment data from these two clinics over an 11-year period (2012–2022), from an ARIA R&V system with treatment plans created in the Eclipse TPS. **Table 1** enumerates the total patients and plans collected during this trial. From both clinics, a total of 6,022 patients, and 13,904 plans, were referenced by treatment records. The framework successfully retrieved 13,871 plans, for a success rate of 99.76%. The small number of plans that were not retrieved were investigated. In many of these cases the RTRECORD referenced a non-patient plan or a plan from several years prior to the designated search window. It took an average of about 18 minutes per patient to process and transfer the data.

The sections below present patient and plan counts based on the earliest Radiotherapy Treatment Record (RTRECORD) and categorized by month and year of treatment.

| Table 1. Summary of data collection volumes from the preliminary testing of the proposed data ingestion framework at two clinics which primarily utilized clinical systems that natively support DICOM Query/Retrieve. | | | |
|---|---|---|---|
|  | Clinic 1 | Clinic 2 | Total |
| **Unique Patients** | 2563 | 3459 | 6022 |
| **Unique Plans Referenced** | 4784 | 9120 | 13904 |
| **Plans Collected** | 4782 | 9089 | 13871 |
| **Success Rate** | 99.95% | 99.66% | 99.76% |

Patient Counts by Clinic

The patient counts per year for both clinics are summarized in **Figure 5.** Clinic 1 exhibited a relatively stable patient volume, with annual counts ranging from approximately 200 to 250 patients across most years. A notable peak occurred in 2017–2018, with patient numbers reaching close to 300, followed by a gradual decline that stabilized around 2021. Clinic 2, in contrast, showed higher patient volumes in the earlier years of the dataset, peaking around 2014 with approximately 450 patients. After this peak, there was a steady decline, with counts reaching around 150–200 patients by 2022. Similar to Clinic 1, all patient data for Clinic 2 included in the analysis were from the ARIA R&V system with Eclipse TPS plans, despite the clinic having other systems that contributed only a minor portion of data, excluded from these figures. Both panels in **Figure 5** reveal seasonal patterns, with noticeable variations in patient counts across months, suggesting cyclical trends in patient intake or treatment schedules. The seasonal trend in plan counts for both clinics parallels the patient count data, underscoring the influence of patient intake cycles on treatment planning activity.

Summary of Collected Data

In addition to patient counts, we collected detailed data on imaging modalities and treatment plans for each clinic. This includes the total number of patients and counts of CT, MR, and PET scans, with a breakdown of CTs referenced by RTSTRUCT, as well as various types of registrations (REG) between imaging modalities, such as CT-CT, CT-MR, and CT-PET. The collected data also categorizes treatment plans by type—curative, palliative, and prophylactic—alongside counts of absent and invalid plans. Invalid plans are further classified into categories such as verification and machine QA. While Clinic 1 exclusively utilized the ARIA R&V system with Eclipse TPS, Clinic 2 employed additional planning systems; however, only data from ARIA/Eclipse were included in the analysis. This comprehensive dataset provides insights into patient volumes, imaging studies, and treatment plan characteristics for each clinic over the study period.

**4. Discussion**

To fully realize the potential of big data in radiotherapy, it is essential to capture both treatment planning and actual treatment delivery information, as insights from correlations between outcomes and treatments depend on this comprehensive data. However, automating the extraction of such data—while preserving the connections between delivery records (RTRECORDs) and planning information (RT Plans, Doses, Structure Sets, Images)—poses significant challenges. Chief among these challenges is the issue of system variety.

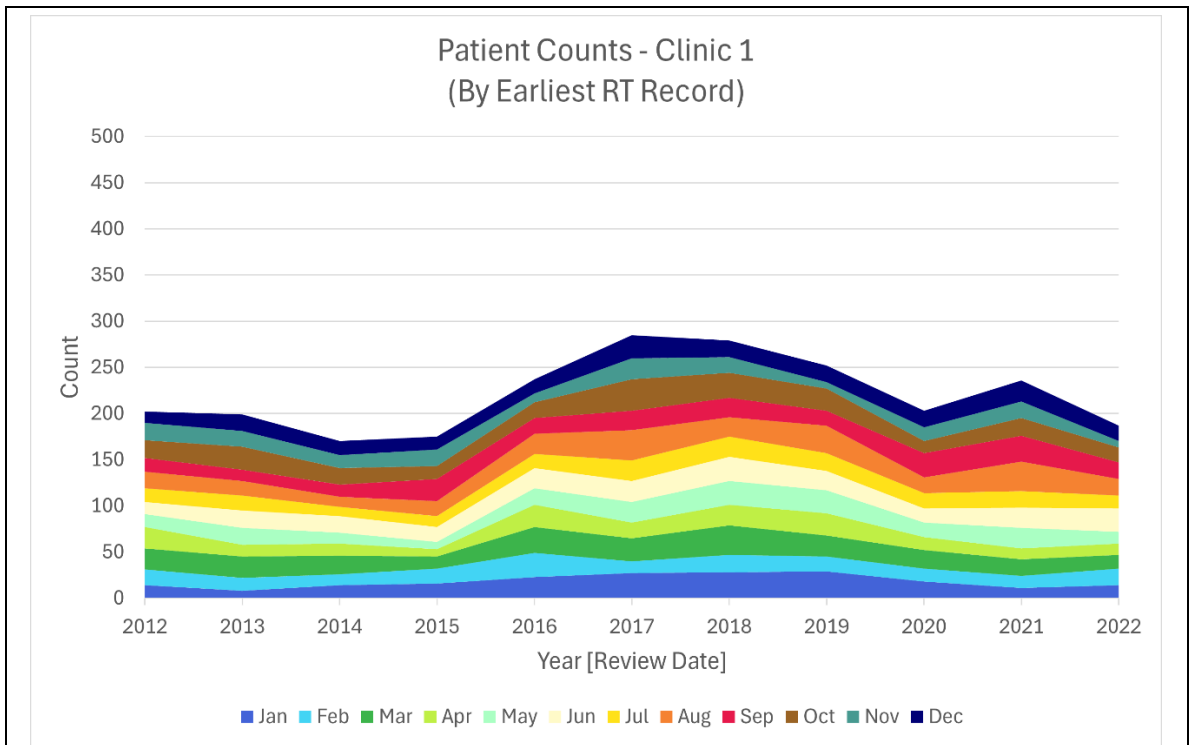

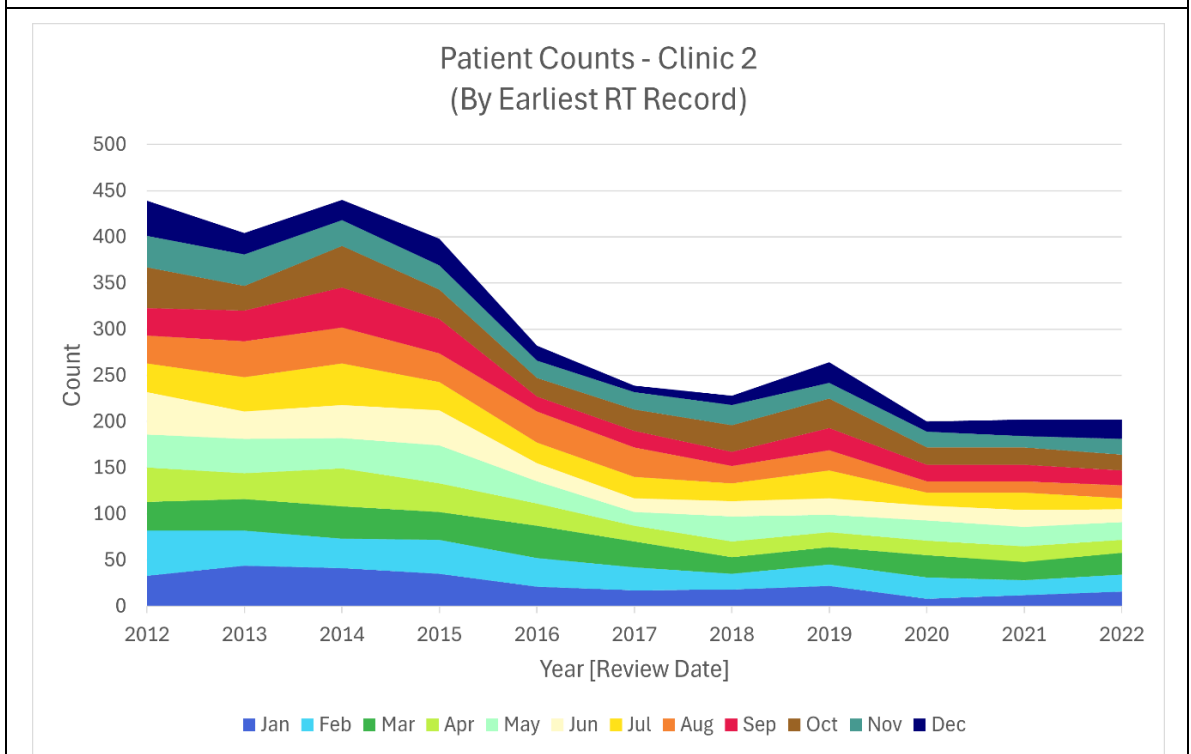

**Figure 5:** Patient counts from 2012 to 2022 for Clinic 1 (top) and Clinic 2 (bottom) categorized by month.

It is common for clinics to use diverse planning systems and technologies, making it difficult to rely on a single data repository for all relevant radiotherapy information, especially when treatment delivery data is included. This diversity, compounded by the unidirectional references in the DICOM standard, increases the complexity of our CORE logic, which must reconstruct these relationships from disparate sources of delivery and planning data. This process is labor-intensive and can sometimes be counterintuitive. For example, we found that one vendor deviated from the common referencing pattern: while most RT Plans reference an RT Structure Set, in TomoTherapy, the RT Dose—not the RT Plan—references the RT Structure Set.

From this experience, we recommend that the DICOM standard adopt reciprocal referencing to enhance data interconnectedness. However, such an improvement would only be effective if clinical data repositories widely support DICOM Query/Retrieve (Q/R). One limitation of our proposed framework is that many clinical systems do not support DICOM Q/R. While the two clinics included in this study used Varian ARIA as their R&V system, which does support DICOM Q/R (though it does not support automated export of DVH data within the DICOM), we recognize that this is not representative of all clinical environments.

To address this limitation, we are currently working on collecting data from two additional clinics that have systems lacking DICOM Q/R capabilities. This effort aims to demonstrate the adaptability of our framework to environments where direct DICOM Q/R is not feasible. For these systems, data are exported through alternative methods (as shown in **Figure 2)**. If the vendor supports it, automated scripts are utilized to extract the data. Otherwise, data are manually exported to an auxiliary PACS, which allows the CORE to link treatment records with planning data despite the absence of DICOM Q/R functionality.

Additionally, some legacy systems at these clinics archive data in proprietary formats, necessitating conversion to DICOM. This conversion is achieved either through custom code or by using an older version of the TPS software provided temporarily by the vendor, enabling us to open and export the data in DICOM format. By extending our data collection methods to these additional clinics, we aim to overcome one of the significant barriers in aggregating radiotherapy data at scale.

Our ongoing work with these clinics highlights the importance of developing flexible data extraction solutions that can accommodate a variety of clinical systems and data formats. It also underscores the need for broader adoption of standardized protocols like DICOM Q/R and for enhancements to the DICOM standard itself. Through this expanded effort, we hope to demonstrate that our framework can be effectively applied across diverse clinical settings, ultimately facilitating more comprehensive and scalable radiotherapy data aggregation.

**5. Conclusions**

Multi-institutional data warehouses will take time to build. While prospective data collection may be more straightforward, to reach a substantive number of cases, retrospective data must be collected as well. Retrospective data ingestion is a complex, laborious task that necessitates automation, and demands several layers of quality control and validation. Our software framework proved effective in unifying radiotherapy planning and delivery data from multiple clinical systems. Critical to this achievement was the ability to meticulously back-trace DICOM relationships, ensuring that only pertinent data were retrieved, thus optimizing data download efficiency. Furthermore, the framework exhibited exceptional resilience in the face of errors, system crashes, and network instabilities – a vital feature considering the prolonged periods required for data collection in each clinic. This robustness underscores the framework's reliability and marks a significant step forward in the field of data consolidation in clinical environments.

**Acknowledgements**

This work was supported in part by the US Department of Veterans Affairs National Radiation Oncology Program.

**Disclaimer:** The contents do not represent the views of the U.S. Department of Veterans Affairs or the US government.